\documentstyle[12pt]{article}
\begin{document}
\def \d {{\rm d}}

\title{Chaotic motion in {\it pp}-wave spacetimes}

\author{
J. Podolsk\'y\thanks{E--mail: {\tt podolsky@mbox.troja.mff.cuni.cz}}
,\  K. Vesel\'y
\\ \\ Department of Theoretical Physics,\\
Faculty of Mathematics and Physics, Charles University,\\
V Hole\v{s}ovi\v{c}k\'ach 2, 180~00 Prague 8, Czech Republic.\\ }

\maketitle

\baselineskip=19pt

\begin{abstract}
We investigate geodesics in non-homogeneous vacuum {\it pp}-wave
solutions and demonstrate their chaotic behavior by rigorous analytic
and numerical methods. For the particular class of solutions
considered, distinct ``outcomes'' (channels to infinity) are
identified, and it is shown that the boundary between different
outcomes has a fractal structure. This seems to be the first example
of chaos in exact radiative spacetimes.
\end{abstract}

\vfil\noindent

\vfil
\eject

\section{Introduction}

Over the past few decades the realization that the behavior of some nonlinear
dynamical systems is extremely sensitive to initial conditions has changed
our view on  time evolution in physics, biology, and even economics.
In the context of general relativity, which is a nonlinear dynamical
theory par excellence, the first systems for which  chaotic behavior
of solutions to the Einstein equations has been recognized and studied
were spatially homogeneous but anisotropic cosmological models.
The ``chaotic cosmology'' programme initiated in 1968 by Misner
directed attention to Bianchi IX  (Mixmaster) models. The oscillatory
dependence of their cosmological scale factors in different spatial
directions on time  has been studied both analytically and numerically
in many works (see relevant contributions in \cite{Hol}, \cite{CL1},
and references therein). For a long time, however, it has been discussed
how conclusively and rigorously the existence of chaos in these
relativistic systems has been demonstrated as gauge
invariant measures of chaotic behavior are required.
Complicated nonlinear effects also occur in systems with
coupled gravitational and scalar or Yang-Mills fields where the number of
degrees of freedom is increased compared to purely gravitational
systems (see for example \cite{Chop}-\cite{BL}).

Another important type of problems providing a nonlinear dynamical system in
the context of general relativity is the study of geodesic motion in
a given spacetime. It is well known that Newtonian many-body
systems are chaotic and it is of primary interest to investigate
similar situations in Einstein's theory. In particular, the chaotic behavior
of geodesics in the relativistic analogue of the two fixed-centres
problem was examined by Contopoulos \cite{Conto1},
Dettmann {\it et al} \cite{DFC} and Yurtsever \cite{Yur}, who investigated
null and timelike geodesics in a spacetime consisting of two (fixed) extreme
Reissner-Nordstr\"om black holes. These results were generalized
by  Cornish and Gibbons \cite{CG} to the Einstein-Maxwell-dilaton
two-centre spacetimes. Bombelli and Calzetta \cite{BoCa} and
Letelier and Vieira  \cite{LeVi} studied chaotic geodesic
motion in perturbed Schwarzschild spacetimes. The chaotic behavior of
a spinning test particle in Schwarzschild spacetime was demonstrated
by Suzuki and Maeda \cite{SuMa}. Karas and Vokrouhlick\'y \cite{KaVo},
Sota {\it et al} \cite{SSM} and Vieira and Letelier \cite{ViLe}, \cite{ViLe2}
studied the behavior of test particles in some static axisymmetric spacetimes.
Chaotic motion has also been demonstrated in (topologically
nontrivial) Robertson-Walker spacetimes by Lockhart {\it et al} \cite{LMP}
and Tomaschitz \cite{Tom}.

In our work \cite{PVcha} we announced that geodesic motion in the
well-known class of {\it pp}-waves, which are vacuum type-{\it N}
solutions representing the simplest exact gravitational wave
spacetimes, is also chaotic. Here we present a detailed analysis.
In the next section we investigate the geodesic equations
for a non-homogeneous case. The chaotic behavior
of timelike, null and spacelike geodesics in these spacetimes
is then established by analytic and numerical fractal methods in
sections~3 and~4, respectively. Some concluding remarks are
given in section~5.

\section{Geodesics in {\it pp}-waves}

The metric of the widely known class of vacuum {\it pp}-waves
\cite{KSMH}, plane-fronted gravitational waves with parallel
rays, can be written in the form
\begin{equation}
\d s^2=2\,\d\zeta \d\bar\zeta-2\,\d u\d v-(f+\bar f)\,\d u^2\ , \label{E1}
\end{equation}
where  $f(u,\zeta)$ is an arbitrary function of the retarded time $u$ and
the complex coordinate $\zeta$ spanning the plane wave surfaces $u=$ const.
The non-trivial curvature tensor components are proportional to $f_{,\zeta\zeta}$
so that (\ref{E1}) represents the Minkowski universe when $f$ is linear
in $\zeta$. The case $f=d(u)\zeta^2$ describes the well-known
plane gravitational wave (the ``homogeneous'' {\it pp}-wave) which
has been thoroughly investigated, see \cite{KSMH} for Refs.
This textbook example of an exact radiative solution has also
been used for the construction of sandwich and impulsive waves
\cite{BPR}-\cite{St}.

Here we study motion in non-homogeneous vacuum
{\it pp}-waves. The geodesic equations for the metric (\ref{E1}) are
\begin{eqnarray}
&&\ddot\zeta + \textstyle{\frac{1}{2}}\bar f_{,\bar\zeta}\, U^2 = 0 \ ,\label{E2}\\
&&\dot u=U=const \ ,\label{E3}\\
&&\ddot v+(f_{,\zeta}\dot\zeta+\bar f_{,\bar\zeta}\dot{\bar\zeta})\, U
 +\textstyle{\frac{1}{2}}(f+\bar f)_{,u}\, U^2=0 \ ,\label{E4}
\end{eqnarray}
where dot denotes $d/d\tau$ with $\tau$ being an affine parameter
along the geodesic. In addition, we also assume a condition normalizing
the tangent to the geodesic such that $U_\mu U^\mu=\epsilon$ where
$\epsilon=-1, 0, +1$ for timelike, null or spacelike geodesics,
respectively. In particular, $\tau$ is a proper time for timelike
geodesic observers. This condition can explicitly be written as
\begin{equation}
\dot v=\frac{1}{U}\left[\dot\zeta \dot{\bar\zeta}
  -\frac{1}{2}(f+\bar f)\,U^2-\frac{\epsilon}{2}\right]\ . \label{E5}
\end{equation}
Here we assume $U\not=0$ (for $U=0$ the geodesic equations can simply be
integrated yielding only trivial null geodesics
$\zeta=\zeta_0$, $u=u_0$, $v=v_1\tau+v_0$
and spacelike geodesics
$\zeta=\textstyle{\frac{1}{\sqrt2}}\exp(\hbox{i}\phi_0)\tau+\zeta_0$,
    $u=u_0$, $v=v_1\tau+v_0$,
where $\zeta_0$, $u_0$, $v_1$, $v_0$ and $\phi_0$ are constants).
By differentiating Eq.~(\ref{E5}) with respect to $\tau$ and using (\ref{E2})
we immediately obtain (\ref{E4}) which can thus be omitted.
It suffices to find solutions of (\ref{E2}) since  $v(\tau)$
can then be obtained by integrating Eq. (\ref{E5}) while $u=U\tau+u_0$.

Now we concentrate on the complex equation (\ref{E2}) which has
the same form for timelike, null and spacelike geodesics.
The first integral of (\ref{E2}) for $f$ independent of $u$ is
\begin{equation}
\dot\zeta\dot{\bar\zeta}+U^2{\cal R}e\,f=2E \ , \label{E6}
\end{equation}
where $E$ is a real constant. Using (\ref{E6}) we can further
simplify (\ref{E5}) into
$v(\tau)=U^{-1}(2E-\epsilon/2)\, \tau
  -2U\int {\cal R}e\,f(\zeta(\tau))\,\d\tau$. Moreover, it
indicates that the Hamiltonian for the system of equations (\ref{E2})
written in two real coordinates $x$ and $y$ introduced
by $\zeta=x+\hbox{i}y$ is
\begin{equation}
H=\textstyle{\frac{1}{2}}\left(p_x^2+p_y^2 \right)+V(x, y)\ , \label{E7}
\end{equation}
with a potential $V(x,y)=\textstyle{\frac{1}{2}}U^2\, {\cal R}e\,f$.
For non-homogeneous {\it pp}-wave spacetimes given by
$f\sim\zeta^n$, $n=3, 4, \cdots$, the corresponding polynomial potential
is called an ``$n$-saddle''.
Its shape for $n=3$ and $n=5$  is shown in Fig. 1.
The metric (\ref{E1}) of these spacetimes is invariant
under rotation $\zeta\to\tilde\zeta=\zeta\exp(\hbox{i}2\pi/n)$
so that for any geodesic there exist other $n-1$ geodesics
differing only by rotations. There is also a symmetry
$\zeta\to\bar\zeta$ corresponding to $y\to-y$.

For simplicity, from now on we shall assume the simplest case
$n=3$. Since any constant multiplicative factor of $V$ can be
removed by a suitable rescaling of the  parameter $\tau$,
we can without loss of generality  consider a potential of the
form
\begin{equation}
V(x, y) =\textstyle{\frac{1}{3}}x^3-xy^2\ , \label{E8}
\end{equation}
called a ``monkey saddle''. Surprisingly, the Hamiltonian
(\ref{E7}), (\ref{E8}) is a special case of famous
H\'enon-Heiles Hamiltonian \cite{HH} which is known to be a ``textbook''
example of a chaotic system. Here, however, quadratic terms in $V$
are missing. This particular case of the H\'enon-Heiles Hamiltonian
has rigorously been investigated by Rod \cite{Rod}.

\section{Analytic demonstration of chaos in {\it pp}-waves}

Rod described in detail the chaotic
\footnote{He called it  ``pathological'' since his paper
\cite{Rod} was written long before the term ``chaos'' came into
vogue.}
behavior of orbits in the Hamiltonian system given by
(\ref{E7}), (\ref{E8}). He concentrated on bounded orbits. These
only appear in the positive energy manifolds $H(x, y, p_x, p_y)=E>0$
for which the level surfaces of the potential (\ref{E8}) have the
form of a ``monkey saddle'' (see Fig. 2). The homogeneity of $V$
guarantees that the orbit structure for any two positive values of
$E$ is isomorphic modulo a constant scale factor and adjustment of time:
$x\to \tilde x= \lambda x$, $y\to \tilde y= \lambda y$ and
$\tau\to \tilde\tau= \tau/\sqrt\lambda$ results in
$E\to \tilde E= \lambda^3 E$.
Therefore, without loss of generality we can restrict attention
to any arbitrary value of $E$, say $E=\frac{1}{3}$. Geodesics for
all other values of $E$ can simply be obtained by rescaling space
coordinates and the affine parameter.

There are three simple geodesics $L_j(\tau)$, $j=1, 2, 3$, for
a given $E$ which will be described in detail in the next section.
In projection to the ($x, y$)-plane they follow the axes $y=0$ and
$y=\pm\sqrt3\,x$ of the three symmetric channels of the connected
region bounded by the level curves $V(x,y)=E>0$ (for notation see
Fig. 2). The boundary consists of three disjoint branches given by
$y=\pm\sqrt{\textstyle{\frac{1}{3}}(x^2-1/x)}$ with asymptotes
$x=0$, $x=\pm\sqrt3\,y$: $V_1$ intersecting the $x$-axis, $V_2$ above
and $V_3$ below the $x$-axis, respectively.

In order to obtain a description of the structure of bounded orbits
Rod first constructed  three basic {\it periodic orbits} $\Pi_j(\tau)$
in each of the three channels (Fig.~2) having perpendicular
intersections with $L_j(\tau)$. Rod proved that these three periodic orbits
are {\it unstable} and in fact are {\it isolated invariant sets}
for the flow generated  by the equations of motion.
He also showed the existence of two additional periodic orbits $\Pi_4$ and
$\Pi_5$ which follow the same trajectory in the ($x, y$)-plane,
only $\Pi_5(\tau)=\Pi_4(-\tau)$.

The analysis of chaotic behavior was then translated in \cite{Rod}
from  the ($x, y$)-plane projection into a topological description of the
{\it asymptotic sets for the periodic orbits} $\Pi_1, \Pi_2, \Pi_3$
in the three-dimensional energy manifold $H=E$. The region in which
these bounded orbits occur can be decomposed
into three cells $R_j$  such that, e.g., $\bar R_1$ is the compact,
connected subset of the ($x, y$)-plane bounded by the lines
$\bar D_1$,  $\bar D_2$ and $\bar \Sigma_1$, see Fig. 2
(here the bar denotes a projection of the corresponding set from the energy
manifold $H(x,y,p_x,p_y)=E$ to the ($x, y$)-plane).
Formally,  $\bar D_1=\{(x, y)\in L_2 \hbox{\ with\ } y\ge0   \}$,
$\bar D_2$ and $\bar D_3$ are defined analogously using the
symmetry and  $\bar\Sigma_1$ denotes the minimum distance line segment
between $V_2$ and $V_3$ (this intersects the branches $V_2$ and $V_3$ in
points given by $y=\pm x$); $\Sigma_2$, $\Sigma_3$ are again defined
analogously. Thus, $\Sigma_j$ and the two $D_k$
are boundaries of the isolating block $R_j$
for the flow containing the unstable periodic solution $\Pi_j$.
Each of the three cells $R_j$ contains one such orbit and
no other bounded orbits; $\Pi_j$ is the only invariant set in $R_j$,
i.e., it is isolated.
Note also that any orbit that leaves the central region
$R_0=R_1\cup R_2\cup R_3$ never enters it again due to the
structure of the acceleration field in the channels beyond
$\Sigma_j$. Such orbits ``go to infinity'' through the
$j$-channel.

Subsequently, Rod presented a description of {\it orbits asymptotic to
the basic periodic orbits} $\Pi_j$ as $\tau\to\pm\infty$. It is appropriate
to  localize these sets of asymptotic orbits by their intersection with
the boundaries $D_k$ which are closed topological two-discs (Fig. 3). Any
point $P$ in the disc represents a point in the phase space section
whose spatial coordinates are all points $(x,y)\in \bar D_k$
while impulses are restricted by the energy condition implying
$p_x^2+p_y^2=r^2\equiv 2[E-V(x,y)]$. The radius $r$
monotonically  increases from $0$ to $\sqrt{2E}$ as the points
$(x,y)\in \bar D_k$ are taken between the intersection of $\bar
D_k$ with $V$ and the origin $(0,0)$. Therefore, we can
identify any point $P=(p_x, p_y)$ of the disc with an orbit
passing through the point $(x(r),y(r))\in \bar D_k$ with a given
velocity $(\dot x, \dot y)=(p_x,p_y)$. Using this we can indicate
in $D_k$ important points, areas and their boundaries relevant
to the chaotic behavior of orbits, see Fig. 3 for $D_3$
(figures for $D_1$ and $D_2$ can simply be obtained by symmetry).
In the figure, any arc $a^+(j)$ denotes
the boundary of points representing orbits that go to $\Sigma_j$
(and then to infinity) as $\tau$ increases; orbits given by
$a^+(j)$ approach $\Pi_j$ as $\tau\to+\infty$. Similarly $a^-(j)$
describes orbits asymptotic to $\Pi_j$ as $\tau\to-\infty$.
Points $l_j$ denote the intersections that
$L_j(\tau)$ has with the circle of velocities at the origin
(orbits pointed towards $l_j$ continue directly to the
$j$-channel as $\tau\to+\infty$).

The analysis in \cite{Rod} showed that the asymptotic sets which
are bordered by $a^\pm(i)$ {\it intersect transversally} on $D_k$
thus defining  open sets $W(m,n)$ called ``windows'' and that
the intersection of the closure of any window with its bounding
sets is {\it uncountable}. This gives
the existence of  orbits that ``connect'' the three unstable
periodic orbits $\Pi_j$: they are {\it homoclinic}
(asymptotic to the same periodic orbit in both time directions)
or {\it heteroclinic} (asymptotic to two different periodic
orbits, one in each time direction). It is the existence of
bounded orbits which temporarily enter the region $R_j$, winding
past $\Pi_j$ and then emerging for recycling through
the next region,  and the existence of homoclinic and heteroclinic
orbits that illustrates complicated chaotic
structure of the flow in the central region $R_0$.

In order to describe the topology of all possible orbits
in phase space it is convenient to introduce
a set of sequences (finite, infinite, or
biinfinite), $s\equiv\cdots, s_k, s_{k+1}, s_{k+2}, \cdots$, where
$s_k\in\{1,2,3\}$, $s_k\not=s_{k+1}$. Any sequence then
corresponds to a sequence $\cdots, R_{s_k}, R_{s_{k+1}}, R_{s_{k+2}}, \cdots$
of blocks through which the orbit is running in the prescribed
order as $\tau$ goes from $-\infty$ to $+\infty$. Using this symbolic
dynamics, Rod defined thirteen disjoint orbit classes intersecting $R_0$:
\begin{enumerate}
\itemsep 1pt
\item  bounded orbits with $s=\{s_k\}_{-\infty}^{-\infty}$
       which do not leave $R_0$
\item  homoclinic orbits with $s=s_1, s_2, \cdots, s_1$
       that come from $\Pi_{s_1}$ and go through the finite sequence to
       the same $\Pi_{s_1}$
\item  heteroclinic orbits with $s=s_1, s_2, \cdots, s_f$
       that come from $\Pi_{s_1}$ and go to different $\Pi_{s_f}$
\item  orbits given by $s=s_1, s_2, \cdots$ that come from $\Pi_{s_1}$
\item  orbits given by $s=\cdots, s_{-2}, s_{-1}$ that go to $\Pi_{s_{-1}}$
\item  orbits that come from $\Pi_{s_1}$ and go through a finite
       sequence  $s=s_1, s_2, \cdots,  s_f$ to $\Sigma_{s_f}$
\item  orbits that come from $\Sigma_{s_b}$ and go through a finite
       sequence  $s=s_b, \cdots,  s_{-2}, s_{-1}$ to $\Pi_{s_{-1}}$
\item  orbits that come from $\Pi_j$ and go to $\Sigma_j$
       intersecting only the region $R_j$ for $j=1,2,3$
\item  orbits that come from $\Sigma_j$ and go to $\Pi_j$
       intersecting only the region $R_j$ for $j=1,2,3$
\item  orbits that come from $\Sigma_{s_1}$ and go through the
       sequence $s=s_1, s_2, \cdots$
\item  orbits that go through the  sequence $s=\cdots, s_{-2}, s_{-1}$
       to $\Sigma_{s_{-1}}$
\item  unbounded  orbits that come from $\Sigma_{s_1}$ through a
       finite sequence $s=s_1, s_2, \cdots, s_f$ to $\Sigma_{s_f}$
\item  orbits that come from $\Sigma_j$ to the same $\Sigma_j$
       intersecting only the region $R_j$ for $j=1,2,3$
\end{enumerate}
Clearly, considering the reversibility of time, the classes
5., 7., 9. and 11. are equivalent to 4., 6., 8. and 10,
respectively. Also, the union of orbit classes 1. through 5.
together with the periodic orbits $\Pi_1, \Pi_2, \Pi_3$ form the
{\it invariant set of bounded orbits}. Orbits that are unbounded
in only one time direction are in classes 6. through 11., and
those that are unbounded in both time directions are in classes
12. and 13.

Finally, the chaotic structure of bounded orbits was analysed in
\cite{Rod} by (a topological version of) the Smale horseshoe map.
With this technique Rod showed that to any biinfinite
sequence of symbols $\{1, 2, 3\}$ there exists an {\it uncountable
number of bounded orbits} running through the blocks $R_j$ in
a given sequence. (Similarly, all the orbit classes 4. through 13.
each contains an uncountable number of orbits.) Also, the flow admits
at least a countable number of homoclinic and heteroclinic orbits.

Rod remarked that the results could be refined if the basic
orbits $\Pi_j$ were known to be hyperbolic so that they would
admit stable and unstable asymptotic manifolds. Consequently,
to each periodic symbol sequence there would correspond a countable
collection of {\it periodic} orbits. The difficulties in proving
the hyperbolicity of $\Pi_j$ were subsequently overcome (in an even
more general context) in \cite{RPC}: the ``monkey saddle''
potential studied above is a special case of ``Example~A'' of
\cite{RPC} given by $V(x,y)=\frac{1}{3}\prod_{i=1}^3(x-\rho_i y)$
with $\rho_1=0$, $\rho_2=\sqrt3$, $\rho_3=-\sqrt3$.

In \cite{CR3}, summarizing and generalizing some previous results
\cite{CR1}, the Hamiltonian (\ref{E7}), (\ref{E8}) (as a particular
case of the H\'enon-Heiles Hamiltonian) was presented as an example
of a system for which the Smale horseshoe mapping can be {\it explicitly}
embedded as a subsystem into the flow along the nondegenerate homoclinic
and heteroclinic orbits to hyperbolic unstable periodic orbits.
The complex behavior of nearby orbits then implied the nonexistence
of global second analytic integral. This completed the proof of
the chaotic nature of the studied system in the sense of a rigorous
definition of chaos, cf. \cite{Chur}.

\section{Numerical demonstration of chaos in {\it pp}-waves}

The equations of motion resulting from (\ref{E7}), (\ref{E8})
have a very simple form
\begin{equation}
\ddot x=y^2-x^2,\qquad  \ddot y=2xy\ .           \label{E10}
\end{equation}
However, their explicit analytic solutions can only be found in very
special cases. Of course, there are (unstable) geodesics with $E=0$
given by $x=0=y$.
Less trivially, letting $y=0$ for all $\tau$, one gets radial geodesics
$L_1(\tau)$ through  the $y=0$ channel (analogous geodesics
$L_2(\tau)$ and $L_3(\tau)$ through remaining two channels can
simply be obtained by rotation, see Fig.~2). Solutions of this type starting
at $\tau=0$ from the level curve $V_1=E$ (i.e., $x(0)=\root3\of{3E}$,
$\dot x(0)=0$) are given by
\begin{equation}
x(\tau)=\root 3 \of{3E}\, \biggl[1-\sqrt{3}\ {1-cn(\beta \tau) \over
1+cn(\beta \tau)} \biggr]\ ,               \label{E11}
\end{equation}
where $\beta^2=3^{-1/6}\,2\,\root3\of{E}$ and
$cn$ is the Jacobian elliptic function with modulus
$k^2=(2+\sqrt3)/4$, i.e., $k=\sin(\frac{5}{12}\pi)$.
The function $x(\tau)$ monotonically decreases from $x(0)=\root3\of{3E}$
across $x(\tau_b)=0$ to $x(\tau_s)=-\infty$ (where the singularity
is located). Interestingly, the
proper times $\tau_b$ and $\tau_s$ are {\it finite}. In fact, one
can easily calculate that $\beta\tau_s=2K(k)\approx5.53612629$
where $K(k)$ is the complete elliptic integral of the
first kind, and    $\beta\tau_b=F(k,\varphi)\approx1.84537543$
where $F(k,\varphi)$ is the elliptic integral of the first
kind with $\sin\varphi=3^{-1/4}2/(1+\sqrt3)$. Clearly,
$\tau_s=3\tau_b$.

In particular, for $E=0$ the solution with $y=0$ is simply
\begin{equation}
x(\tau)=-{6 \over (\tau-\tau_s )^2}\ ,               \label{E12}
\end{equation}
which describes (for $\tau \geq \tau_s$) a geodesic emerging from
the singularity
$x=-\infty$ at $\tau=\tau_s$ and approaching the origin $x=0$
asymptotically as $\tau\to\infty$, or (for $\tau \leq \tau_s$)
a geodesic that falls from the origin  to the singularity.
Again, if we set the initial condition $x(0)=x_0<0$ then
$\tau_s=\sqrt{-6/x_0}$ so that the time needed to fall from
$x_0$ to the singularity is finite.

We also considered to localize the {\it trajectories} of the basic
periodic orbits $\Pi_j$. Due to the symmetry one can concentrate
on $\Pi_1$ only which can be described as a function $x(y)$. Assuming
$E=\frac{1}{3}$, it must be a solution of a non-linear equation
\begin{equation}
\frac{1+3xy^2-x^3}{1+x'^2}x''+3xyx'={\textstyle\frac{3}{2}}(y^2-x^2)
\ ,   \label{E12a}
\end{equation}
where $x'=\d x/\d y$, such that $x(-y)=x(y)$ and $x'(0)=0$.
Then  $\Pi_1$ is given by
\begin{equation}
x=a+by^2+cy^4+dy^6+\cdots \ ,               \label{E12b}
\end{equation}
where $b=\frac{3}{4}a^2/(a^3-1)$,
$c=\frac{1}{16}(2+13a^3)/(a^3-1)^2$,
$d=\frac{1}{320}a(505a^6-437a^3-68)/(a^3-1)^4$.
We found the value of $a$ numerically and then calculated
the remaining constants using the above relations:
$a=-0.5152$, $b=-0.1751$, $c=0.0107$, $d=-0.0012$.

The explicit geodesics presented above are very special. In order to
get the global picture of a motion one has to perform the
integration of Eqs. (\ref{E10}) numerically. Typical geodesics in
the studied spacetime are shown in Figs.~4~and~5.

In Fig.~4 we present geodesics starting from the branch $V_2$.
The curves are orthogonal to $V_2$, proceed first downwards
and ``fan out'' from both sides of $\Pi_1$ (and also $\Pi_3$).
Such behavior illustrates some of the analytic results presented
in \cite{Rod} and indicates that all $\Pi_j$ are unstable.

Other geodesics that we obtained by numerical integrations are shown in
Fig.~5. Their initial conditions are chosen such that
the geodesics start at $\tau=0$ from a circle $x^2+y^2=\rho^2$ in the
($x, y$)-plane. Due to a scaling property of the ``monkey
saddle'' potential (see section~3) we can, without loss of
generality, assume $\rho=1$ (all other geodesics except those
intersecting the origin can simply be obtained by rescaling).
In Fig.~5a we present geodesics
with $\dot x=0=\dot y$ at $\tau =0$ and in Fig.~5b geodesics
starting with non-vanishing (but same) velocities.
(Note that these conditions are different from the approach adopted
in the previous section since the geodesics are not on the {\it same}
``energy manifolds'' $E=$ const. However, $E$ {\it is not}
the energy of particles and photons and there is no physical reason
to sort all the geodesics according to $E$ no matter how useful
it proved to be from the mathematical point of view).

We observe that each {\it unbounded} geodesic escapes
to infinity (i.e., falls to the singularity)
through {\it only one} of the three channels in the potential.
Choosing non-zero initial velocities
makes more geodesics prefer one of the channels (cf. Fig.~5a
and Fig.~5b) but does not significantly change the character of
the motion.

In fact, all unbounded geodesics through the three
channels oscillate around the corresponding ``basic'' radial
geodesics $L_j(\tau)$ discussed above.  Let us assume geodesics
through the first channel centered by $L_1(\tau)$ given by (\ref{E11})
(similar results for the second and the third channel follow by
symmetry). As they approach the singularity at $x=-\infty$, we
may assume $|y|\ll|x|$ (this is also justified  by our numerical
simulations). Then the asymptotic solution to Eq. (\ref{E10}) is
$x(\tau)\approx-6(\tau_s-\tau)^{-2}$ and therefore
\begin{equation}
y(\tau)\approx\sqrt{\tau_s-\tau}\left(A\cos\left[
\textstyle{\frac{\sqrt{47}}{2}}\ln(\tau_s-\tau) \right]
+B\sin\left[\textstyle{\frac{\sqrt{47}}{2}}\ln(\tau_s-\tau)\right]\right)
\ ,               \label{E13}
\end{equation}
where $A$ and $B$ are arbitrary constants. As the
geodesics approach the singularity, their frequency of
oscillations around $L_1(\tau)$ grows to infinity while the
amplitude of  oscillations tend to zero. We call this effect
a ``focusation''.

The main objective of this section, however, is to establish the
chaotic behavior of geodesics in {\it pp}-waves  by numerical
means. This may be more illustrative than the formal analysis
presented in the previous section. Chaos is usually indicated by a
sensitive dependence of possible outcomes on the  choice of
initial conditions. The standard approach, called a fractal
method, was advanced in the papers by Cornish, Dettmann, Frankel,
Levin and others (see for example \cite{CL1},
\cite{Conto1}-\cite{CG}). It starts with a definition of several
different outcomes, i.e., types of ends of all possible
trajectories. Subsequently, a set of initial conditions is
evolved numerically until one of the outcome states is
reached. Chaos is established if the basin boundaries which separate
initial conditions leading to different outcomes are fractal.
As we shall now demonstrate, we observe exactly these structures
in the studied system.

It is natural to parametrize the unit circle from which the
geodesics in Fig~5a start at $\tau=0$ (with vanishing velocities)
by $x(0)=\cos\phi, y(0)=\sin\phi$, $\phi\in [-\pi,\pi)$.
We have already pointed out that all unbounded geodesics
approach the singularity at infinite values of $x$ and $y$
through only {\it three distinct channels}. These represent
possible outcomes of our system and we assign them  symbol
$j$ which takes one of the corresponding values, $j\in\{1, 2, 3\}$
(thus, for example, $j=1$ means that the geodesic approaches the
infinity at $x=-\infty$, $y=0$ through the first channel centered
by $L_1(\tau)$ as $\tau\to\tau_s>0$). From Fig~5a we observe that the
behavior of the function $j(\phi)$ depends very sensitively on $\phi$
in certain regions. We calculated $j(\phi)$ numerically
and we display the results in Fig~6
\footnote{When the values of $j$ were colour coded we obtained
nice fractal pictures. Unfortunately, here we could present only
their black-and-white versions which are not so impressive.
Therefore, we chose simply to plot the function $j(\phi)$.}.
Also, we plot in the same diagram the function $\tau_s(\phi)$
which takes the value of the parameter $\tau$ when the singularity
is reached by a given geodesic.

The boundaries between the outcomes appear to be fractal.
This is confirmed on the enlarged detail of the image and the
enlarged detail of the detail etc. up to the sixth level.
In Fig.~7 we show such zooming in of the fractal interval localized around the
value $\phi\approx0$ (there are two similar fractal
intervals  around $\phi\approx\frac{2}{3}\pi$ and
$\phi\approx-\frac{2}{3}\pi$ corresponding to the other outcome channels).
On {\it each level} the structure has the same property, namely that between two larger
connected sets of geodesics with outcome channels $j_1$ and $j_2\not=j_1$
there is always a smaller connected set of geodesics with outcome
channel $j_3$ such that $j_3\not=j_1$ and $j_3\not=j_2$.
Similarly as in \cite{Conto1}, \cite{CG}, the structure of initial conditions
of these three types of orbits resembles three mixed Cantor sets,
and this fact is again a manifestation of chaos.

The above structure of $j(\phi)$ has its counterpart in
the fractal structure of $\tau_s(\phi)$, see Fig.~6 and Fig.~7.
We observe that the value of this function increases considerably
on each discontinuity of $j(\phi)$, i.e., on any fractal basin boundary
between the different outcomes (in fact, $\tau_s$ is {\it infinite}
there). Thus, there is an infinite number of peaks, each of which
corresponds to an orbit asymptotic to one of the periodic
orbits $\Pi_j$ as $\tau_s\to\infty$ (these orbits never
``decide'' on a particular outcome, and so never escape to infinity).
Also, the value of $\tau_s$ increases in the non-chaotic regions
of $\phi$ as one zooms into the higher levels of the fractal.
This can be understood physically. Most geodesics fall into the
singularity {\it directly}. In Fig.~6 they form the three largest
connected sets of length $\Delta\phi_0\approx1.83$ on which $j(\phi)$
is constant. Three smaller connected sets of length
$\Delta\phi_1\approx0.248$ in the first level of the
fractal structure correspond to geodesics that approach the
singularity {\it after one ``bounce''} on the potential wall
$V_j$ in the central region. Their values of $\tau_s$ must naturally
be larger. (Note that they also contain geodesics $L_j(\tau)$
given by $\phi=0$, $\phi=\frac{2}{3}\pi$ and $\phi=-\frac{2}{3}\pi$
for $j=$ 1, 2 and 3, respectively; we  calculated explicitly their
times $\tau_s=2K(k)\approx5.5361/\beta\approx5.1519$ which agrees
with the values in Fig.~6).
In the second level of the fractal there are sets of length
$\Delta\phi_2\approx8.6\times10^{-3}$ with
geodesics that reach the singularity {\it after two ``bounces''}
having even larger values of $\tau_s$ etc. We found numerically that
$\Delta\phi_3\approx3.23\times10^{-4}$,
$\Delta\phi_4\approx1.18\times10^{-5}$,
$\Delta\phi_5\approx4.28\times10^{-7}$,
$\Delta\phi_6\approx1.56\times10^{-8}$, etc.

Using this we can estimate the fractal dimension $D$ of the above
structure. We observe that $\Delta\phi_i/\Delta\phi_{i+1}\to r=27.4\pm0.2$
 for large values of $i$. Since the ``fractal
pattern'' is doubled in each step we get
$D=\ln2/\ln r=0.209\pm0.001$ so that the dimension is clearly non-integer.

The motion can also be visualized by the time evolution of a ring
of free test particles in the ($x, y$)-plane which are in rest initially.
In Fig.~8 we observe that the circle is deformed in a complicated way.
In fact, the circle forms loops that escape to
{\it different} outcome channels as $\tau$ grows. They approach
the singularity and one can easily check that as the particles
forming the original circle move
in different channels, their relative proper distance grows.
Of course, there are also isolated particles of
the initial ring that will forever remain near the origin
approaching asymptotically the basic periodic orbits $\Pi_j$.

Finally, in order to confirm once more the chaotic behavior of geodesics in
{\it pp}-waves we return back to the Rod analysis. We
checked his main result concerning the motion in the energy
manifold $E=$~const. by numerical simulation of the disc $D_3$.
This is shown in Fig.~9 which significantly improves Rod's schematic sketch
(cf. Fig.~3). There are two narrow chaotic bands with a fractal
structure. Each boundary in the band between two different
outcomes represents some bounded geodesic as $\tau\to\infty$.
Clearly, there is an uncountable number of such geodesics.
A similar set for $\tau\to-\infty$ can simply be obtained by
a reflection with respect to the $y$-axis. An interesting feature of
the disc is that the fractal bands bend as they
approach the boundary of the disc which represents the geodesics
crossing the origin $x=0$ and $y=0$. In Fig.~10 we present some
geodesics of this type. Again, the outcome is extremely sensitive
on initial conditions. It also demonstrates that the higher the
level of the fractal, the greater is the number of ``bounces''
that a geodesic undergoes before falling into the singularity.

\section{Final remarks}

In this paper we have demonstrated by invariant analytic and numerical
methods the chaotic behavior of geodesic motion in non-homogeneous
vacuum {\it pp}-wave spacetimes. It was established for all types
of geodesics: timelike, null and spacelike.  This seems to be the
first explicit demonstration of chaos in {\it exact} radiative spacetimes
(note that a chaotic interaction of particles with particular classes
of linearized gravitational waves on given backgrounds has been
studied in \cite{BoCa}, \cite{LeVi}, \cite{VaPa}-\cite{Koku}).
In fact, {\it pp}-wave solutions represent the simplest models of exact
gravitational waves in general relativity. However, so far most work
has concentrated on homogeneous {\it pp}-waves for which the motions
are integrable (Eq. (\ref{E2}) is linear) and hence non-chaotic.

Although we investigated in detail only the spacetime (\ref{E1})
with the function $f(\zeta)$ of the form $f\sim\zeta^3$
corresponding to the ``monkey saddle'' potential $V$ given by
(\ref{E8}), as we indicated in \cite{PVcha} the results might
be carried out to $f\sim\zeta^n$ with $n\ge4$, i.e., to the general
$n$-saddle potential $V=\textstyle{\frac{1}{n}}\, {\cal R}e\,\zeta^n$
(cf. Fig.~1 for $n=5$). In particular, the decomposition
of the central region into topological isolating
blocks and the existence of basic unstable periodic solutions
$\Pi_j$ in each channel $j=1, 2, \cdots, n$ is immediate \cite{Rod}.
These periodic solutions are hyperbolic \cite{RPC} and the existence of
nondegenerate homoclinic and heteroclinic orbits is established
\cite{CR3}, \cite{CR1}. As an example we present our numerical results
for $n=5$: Fig.~11a shows unbounded geodesics starting from a unit circle
from rest and  Fig.~11b displays corresponding functions $j(\phi)$
and $\tau_s(\phi)$. There are now five outcome channels and again,
the structure is fractal.

Our work demonstrates that geodesic motion in spacetimes
even as simple and well-known as {\it pp}-waves can be complex.
Hopefully, it will initiate investigation of chaos in other exact
radiative space-times.

\section*{Acknowledgments}

We acknowledge the support of grants No. GACR-202/96/0206 and
No. GAUK-230/1996 from the Czech Republic and Charles University.
We thank the developers of the software system FAMULUS
(Tom\'a\v s~Ledvinka in particular) which we used for
computation and drawing of all the pictures. Also, we thank Jerry
Griffiths for his help with the manuscript.

\newpage

\section*{Figure Captions}

\begin{description}

\item{Fig.~1\ } The shape of the potential $V(x,y)=\textstyle{\frac{1}{n}}
\, {\cal R}e\,\zeta^n$, where $\zeta=x+\hbox{i}y$, describing geodesics
in corresponding non-homogeneous vacuum {\it pp}-wave spacetimes
(here for $n=3$ and $n=5$).

\item{Fig.~2\ } For $n=3$, the ``monkey saddle'' potential $V(x, y)
=\textstyle{\frac{1}{3}}x^3-xy^2$ plays a crucial role in famous
H\'enon-Heiles chaotic system. The level surface  $V=E$ consists of disjoint
branches $V_j$, $j=1,2,3$, defining three channels to infinity, each
centered by a special geodesic $L_j(\tau)$. The central region can
be decomposed into three cells $R_j$ bounded by $D_k$ and
$\Sigma_j$ such that the only bounded orbit in $R_j$ is an
unstable periodic orbit $\Pi_j(\tau)$. Also, the trajectory of
two additional periodic orbits $\Pi_5(\tau)=\Pi_4(-\tau)$ which we
found by numerical simulations is drawn by a dashed line.
See the text for more details.

\item{Fig.~3\ } The boundary $D_3$ between the isolating blocks
$R_2$ and $R_3$ is a closed topological two-disc. Any point $(p_x, p_y)$
in this schematic figure represents an orbit passing with a velocity
$(\dot x, \dot y)=(p_x,p_y)$ through the point $(x(r),0)\in \bar D_3$
such that (for $E=\textstyle{\frac{1}{3}}$)
$x=\sqrt[3]{1-\textstyle{\frac{3}{2}}r^2}$ where $r=\sqrt{p_x^2+p_y^2}$.
Arc boundaries of orbits that go to infinity through the $j$-channel
as $\tau\to\pm\infty$ are denoted by $a^\pm(j)$. Their
intersections form ``windows'' $W(m,n)$ containing homoclinic and
heteroclinic orbits asymptotic to and from the basic periodic orbits
$\Pi$.

\item{Fig.~4\ } Geodesics starting from the branch $V_2$
``fan out'' from both sides of $\Pi_1$ and  $\Pi_3$.

\item{Fig.~5\ } Geodesics starting from a unit circle in the
($x, y$)-plane a) from rest, b) with initial velocity
$\dot x=0.4$, $\dot y=0.8$. They escape to infinity (where the
singularity is localized) through only three channels
in the potential and a sensitive dependence of these three possible
outcomes $j\in\{1, 2, 3\}$ on the  choice of initial conditions is observed.

\item{Fig.~6\ } Plot of functions $j(\phi)$ and $\tau_s(\phi)$
which labels the three possible outcomes and the value of $\tau$
when the singularity is reached by a given geodesic, respectively,
on $\phi\in[-\pi,\pi)$ parametrizing the initial position
on a unit circle, $x(0)=\cos\phi, y(0)=\sin\phi$;
$\dot x(0)=0=\dot y(0)$. Boundaries separating different outcomes
are fractal establishing chaos.
Each peak in $\tau_s(\phi)$ which coincides with discontinuity in $j(\phi)$
corresponds to some bounded orbit.

\item{Fig.~7\ } The fractal structure of $j(\phi)$ and $\tau_s(\phi)$
was  confirmed by zooming in the interval around the value $\phi\approx0$
up to the sixth level. Between any two connected sets representing
geodesics with outcome channels $j_1$ and $j_2\not=j_1$ there is always
a smaller connected set with $j_3\not=j_1$ and $j_3\not=j_2$.

\item{Fig.~8\ } The time evolution of a ring of free test particles in
the ($x, y$)-plane, initially in rest. The circle is deformed
in a complicated fractal way with different
segments moving to different outcome channels. Notice, for
example, a similarity of the patterns at $\tau=4.0$ and $\tau=7.0$.
However, at  $\tau=7.0$ the lines are in fact doubled and consist
of particles coming from different parts of the original circle.

\item{Fig.~9\ } Exact form of the disc $D_3$ which we obtained by
numerical simulation significantly improving Rod's schematic sketch
presented in Fig.~3. The chaotic bands in the dics of radius
$\sqrt{\frac{2}{3}}\approx0.8165$ are very narrow and also
have fractal structure which is clear from the enlarged details.
The coding is such that white colour denotes the outcome
channel $j=1$, black corresponds to $j=2$ and grey corresponds to $j=3$.
Each boundary in the fractal basin represents a geodesic
asymptotic to some $\Pi_j$ as $\tau\to+\infty$.

\item{Fig.~10\ } Typical geodesics starting from the origin
$x(0)=0=y(0)$ with velocities
$\dot x(0)=\sqrt{\frac{2}{3}}\cos\psi$,
$\dot y(0)=\sqrt{\frac{2}{3}}\sin\psi$ for
a) $\psi\in(-0.28, +0.28)$, b) $\psi\in(0.024, 0.028)$,
c) $\psi\in(0.0254, 0.0256)$.
Again, the outcome is extremely sensitive on initial conditions.

\item{Fig.~11\ } Motion in {\it pp}-waves given by the structural
function $f\sim\zeta^5$: a) geodesics starting from a unit circle
from rest approach infinity through five outcome channels,
b) corresponding functions $j(\phi)$ and $\tau_s(\phi)$
also have fractal structure.

\end{description}

\end{document}